\documentclass[aps,prd,preprint,superscriptaddress,tightenlines,nofootinbib,showpacs]{revtex4}

\usepackage{graphicx}
\usepackage{dcolumn}
\usepackage{bm}

\def\Y{\ifmmode \Upsilon \else%
$\Upsilon$ %
\fi}
\def\chib{\ifmmode \chi_b \else%
$\chi_b$ %
\fi}
\def\chibp{\ifmmode \chi_b' \else%
$\chi_b$ %
\fi}
\def\Q#1#2#3#4{\ifmmode
 \,#1\,{^{#2}#3}_{#4}
\else%
$#1\,{^{#2}#3}_{#4}$ %
\fi}
\def\eonem#1#2{\ifmmode
\left| <#1|r|#2> \right|
\else%
$\left| <#1|r|#2> \right|$
\fi}

\def\ee{\ifmmode e^+e^- \else $e^+e^-$  \fi}
\def\mm{\ifmmode \mu^+\mu^- \else $\mu^+\mu^-$  \fi}
\def\LL{\ifmmode l^+l^- \else $l^+l^-$  \fi}

\def\etal{{\it et al.}}

\begin{document}

\preprint{CLNS 04/1897}       
\preprint{CLEO 04-16}         

\title{Photon Transitions in $\Upsilon(2S)$ and $\Upsilon(3S)$ Decays}

\author{M.~Artuso}
\author{C.~Boulahouache}
\author{S.~Blusk}
\author{J.~Butt}
\author{E.~Dambasuren}
\author{O.~Dorjkhaidav}
\author{J.~Li}
\author{N.~Menaa}
\author{R.~Mountain}
\author{H.~Muramatsu}
\author{R.~Nandakumar}
\author{R.~Redjimi}
\author{R.~Sia}
\author{T.~Skwarnicki}
\author{S.~Stone}
\author{J.~C.~Wang}
\author{K.~Zhang}
\affiliation{Syracuse University, Syracuse, New York 13244}
% ...
\collaboration{CLEO Collaboration} 
\noaffiliation


\date{10/25/04}

\begin{abstract} 
We have studied 
the inclusive photon spectra in $\Upsilon(2S)$ and $\Upsilon(3S)$ decays 
using a large statistics data sample 
obtained with the CLEO III detector.
We present the most precise measurements of 
electric dipole (E1) photon transition rates and photon energies for
$\Upsilon(2S)\to\gamma\chi_{bJ}(1P)$ and
$\Upsilon(3S)\to\gamma\chi_{bJ}(2P)$ ($J=0,1,2$).
We measure the rate for a rare E1 transition
$\Upsilon(3S)\to\gamma\chi_{b0}(1P)$
for the first time.
We also set upper limits on the 
rates for the hindered magnetic dipole (M1) transitions
to the $\eta_b(1S)$ and $\eta_b(2S)$ states. 
\end{abstract}

\maketitle

Long-lived $b\bar b$ states are especially well suited for
testing lattice QCD  calculations \cite{LatticeQCD}
and effective theories
of strong interactions, 
such as potential models \cite{PotentialModels}.
The narrow triplet-$S$ states, $\Y(1S)$, $\Y(2S)$ and
$\Y(3S)$, are directly formed in $e^+e^-$ collisions.
Six triplet-$P$ states, $\chi_b(2P_J)$ and $\chi_b(1P_J)$ with
$J=2,1,0$, are reached via E1 photon transitions from 
the $\Y(2S)$ and $\Y(3S)$ states.
Measurements of the photon energies determine 
the $P-$state masses, while 
measurements of the transition rates test theoretical
predictions for E1 matrix elements.
Such photon transitions were previously studied by
CUSB \cite{CUSBI}, CUSB II \cite{CUSBII}, 
Crystal Ball \cite{CB}, ARGUS \cite{ARGUS}, 
CLEO \cite{CLEOI} and CLEO II \cite{CLEOII} detectors.
The CLEO III experiment accumulated 
$9.3\times10^6$ $\Y(2S)$ and 
$5.9\times10^6$ $\Y(3S)$ resonant decays,
which constitute an order of magnitude increase in statistics
over previous experiments. 
In this paper, we investigate inclusive photon spectra
in decays of these resonances.
In addition to the study of E1 photon transitions to
$P-$states, we also search for M1 photon
transitions to yet unobserved singlet states,
$\eta_b(1S)$ and $\eta_b(2S)$. 
A similar study of photon transitions
in $\psi(2S)$ decays with the same detector
has been recently reported elsewhere \cite{psipprl}.

The CLEO III detector is equipped with
a CsI(Tl) calorimeter, first
installed in the CLEO II detector \cite{CLEOIIdetector},
with energy resolution matching that
of the Crystal Ball (NaI(Tl) crystals) \cite{CB} 
and CUSB II (BGO crystals) detectors \cite{CUSBII}.  
The finer segmentation of the CLEO calorimeter provides
for better photon detection efficiency and more effective
suppression of the photon background from $\pi^0$ decays
than the previous experiments.
The CLEO III tracking detector, 
consisting of 
a silicon strip detector and a large drift chamber \cite{CLEOIIIDR}, 
provides improved suppression of backgrounds 
from charged particles.
The magnetic field inside the tracking detector was 1.5 T. 

The data used in this analysis were collected at the
CESR $e^+e^-$ storage ring at and near the $\Y(1S)$, $\Y(2S)$
and $\Y(3S)$ resonances.
The $\Y(1S)$ data and the data taken at the continuum below each
resonance are used for background subtraction as described below.
Integrated luminosities accumulated on-(off-)resonance are
$1.06$ ($0.19$), $1.31$ ($0.44$) and $1.39$ ($0.16$) fb$^{-1}$
for $\Y(1S)$, $\Y(2S)$ and $\Y(3S)$, respectively.

\def\ntk{N_{ch}}
The data analysis starts with the selection of 
hadronic events.
We require that the observed number of charged tracks ($\ntk$) 
be at least three.
The visible energy of tracks and photons 
($E_{vis}$) must be 
at least 20\%\ 
of the center-of-mass energy ($E_{CM}$).
For $3\le\ntk\le4$ the total energy visible in the 
calorimeter alone ($E_{cal}$) must be at least 15\% of 
$E_{CM}$ and, to suppress residual $\ee\to\ee(\gamma)$ events, the 
most energetic shower in the calorimeter must be less
than 75\%\ of the beam energy or $E_{cal}<0.85\,E_{CM}$. 
The resulting event selection 
efficiency is $92\%$ for decays of 
the $\Upsilon(nS)$ resonances.

To determine the number of produced resonant decays
we subtract scaled continuum background from 
the number of hadronic events observed in the
on-resonance data and correct for the selection
efficiency. Cosmic-ray, beam-gas and beam-wall
backgrounds constitute less than a percent of
the on-resonance data. They largely cancel in
the subtraction of the off-resonance data. 
The systematic error on the number of 
resonant decays ($2\%$) is dominated by 
uncertainty in Monte Carlo modeling of
hadronic annihilation of the $b\bar b$ states.

In the next step of the data analysis we look for photon candidates
in the selected hadronic events.
Showers in the calorimeter are required not to match the
projected trajectory of any charged particle, and
to have a lateral shower profile consistent with
that of an isolated electromagnetic shower.
We restrict the photon candidates to 
be within the central barrel part of the calorimeter
($|\cos\theta|<0.8$) where the photon energy resolution is
optimal. 
The main photon background in this analysis comes from
$\pi^0$ decays. We can reduce this background by removing 
photon candidates that combine with another photon 
to fit the $\pi^0$ mass.
Unfortunately, this lowers the signal efficiency, 
since random photon combinations sometimes fall within the 
$\pi^0$ mass window.
Our studies show that there is no benefit to applying the $\pi^0$
suppression for photon energies around 100 MeV where
many of the dominant E1 photon lines are observed.
This is in contrast with the recently analyzed 
$\psi(2S)$ decays \cite{psipprl} and is caused by 
the higher shower multiplicity in $b\bar b$ decays.
The situation is different for higher photon energies.
Two photons in a decay of a fast $\pi^0$ must be spatially close
to each other, which decreases the rate of false $\pi^0$ 
candidates.
We suppress high energy photons, 
which match the $\pi^0$
mass when combined with another photon which satisfies 
$\cos\theta_{\gamma\gamma}>0.7$, where $\theta_{\gamma\gamma}$
is the opening angle between the two photons.

\begin{figure}[htbp]
\includegraphics[width=5in]{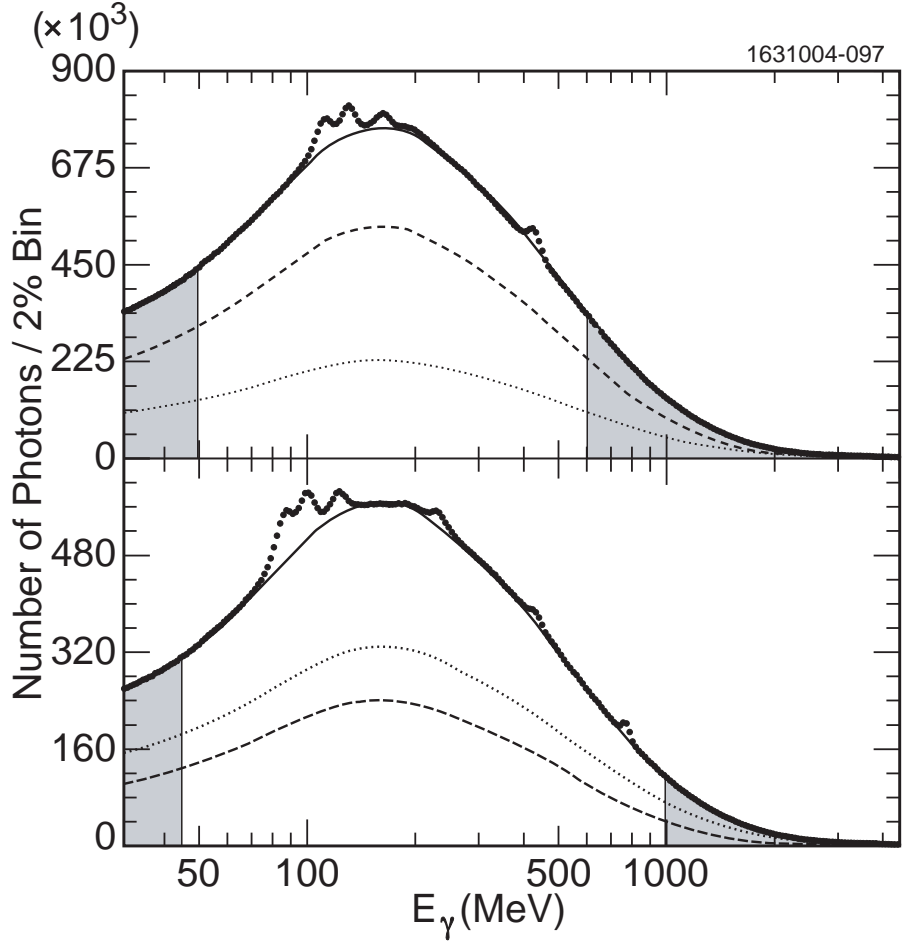}
\caption{
Fit of the off-resonance and $\Y(1S)$ photon spectra to
photon backgrounds in the $\Y(2S)$ (top) and $\Y(3S)$ (bottom) data.
The energy regions used in the fit are shaded.
The total fitted background is represented by the solid line.
The $\Y(2S)$ and $\Y(3S)$ data are shown by points.
The fitted contributions of the off-resonance (dashed line)
and $\Y(1S)$ spectra (dotted line) are also shown.
See the text for explanation of various photon lines
observed in the data.
\label{fig:all}
}
\end{figure}

Photon energy spectra obtained without $\pi^0$ suppression
are shown for $\Y(2S)$ and $\Y(3S)$ decays in Fig.~\ref{fig:all}.
Unlike in the $\psi(2S)$ photon spectrum, even dominant E1 peaks
from $n^3S_1\to\!(n\!-\!1\!)^3P_J$ transitions have a small
signal-to-background ratio. This is not only due to the increased
shower multiplicity but also to a significant continuum background
in the on-resonance data. Furthermore,
because the $b\bar b$ system is less relativistic, 
the three lines are more closely spaced due to the smaller
fine-structure mass splitting.
Therefore, estimation of the background level under the peaks
is more challenging in this analysis. 
To constrain the background shape under the peaks we
use the photon spectrum observed in the off-resonance data and
in the $\Y(1S)$ sample. 
The fraction of each is determined by a fit to the
$\Y(2S)$ or $\Y(3S)$ photon spectra
in the energy range free of any photon peaks.
These regions are shown as shaded areas in Fig.~\ref{fig:all}.
This procedure gives us a good approximation for the shape of 
these rapidly varying background components as
illustrated in Fig.~\ref{fig:all}.
Since photon backgrounds in decays of the $\chi_b$ states
may deviate somewhat from the $\Y(1S)$ photon spectrum, 
we allow for an additional, 
slowly varying background component, represented by
a low order polynomial when fitting the peak region.
The overall normalization of the continuum+$\Y(1S)$ background
is allowed to float in this fit. Systematic errors due to
the background parametrization are evaluated by varying
the fit ranges and the order of the polynomial.

\begin{figure}[htbp]
\includegraphics[width=5in]{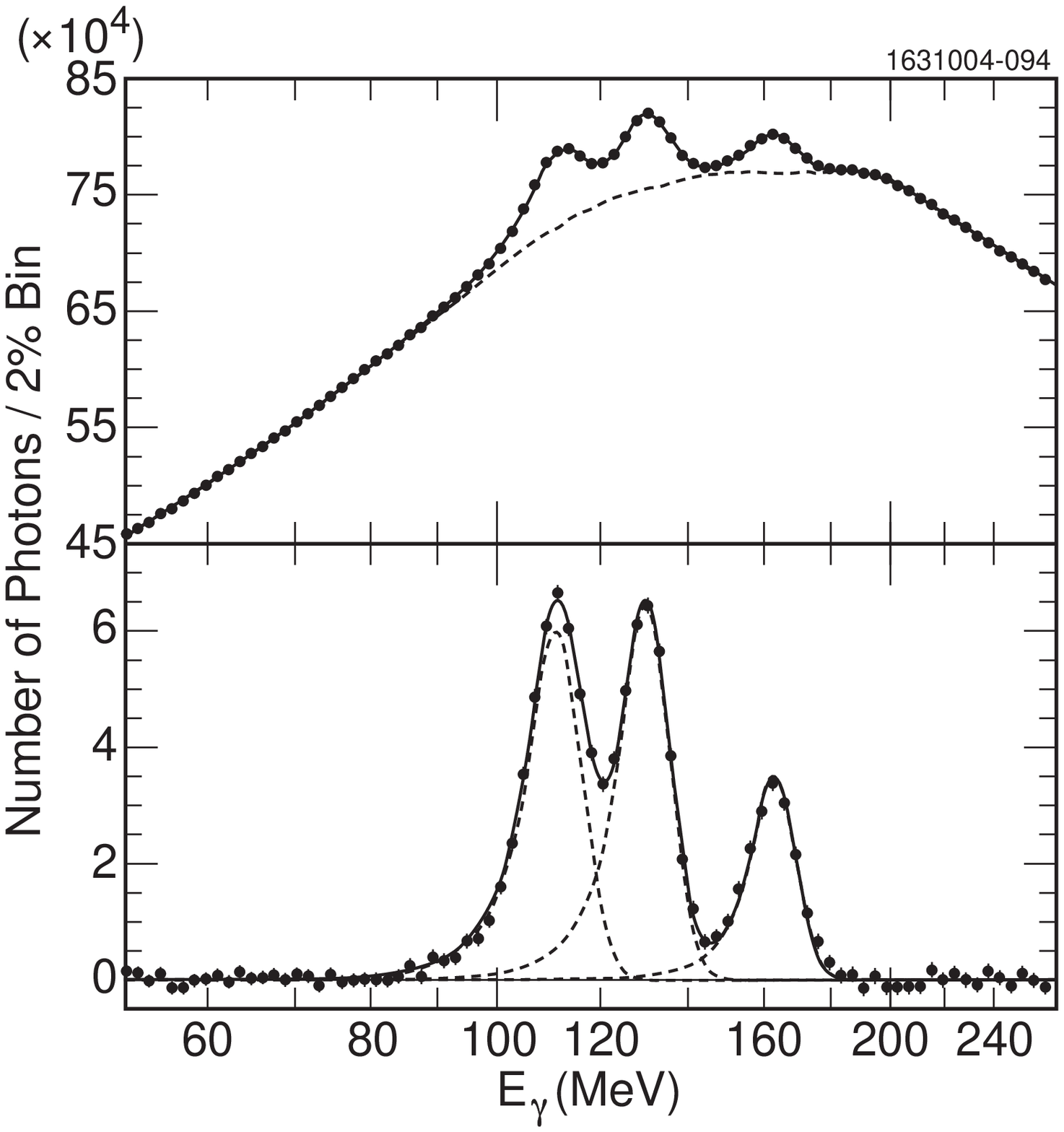}
\caption{
\label{fig:2sfit}
Fit to the $\Y(2S)\to\gamma\chi_{bJ}(1P)$ ($J=2,1,0$) photon
lines in the data.
The points represent the data (top plot).
Statistical errors on the data 
are smaller than the point size.
The solid line represents the fit.
The dashed line represents total fitted
background. 
The background subtracted data 
(points with error bars) are shown 
at the bottom.
The solid line represents the fitted
photon lines together. The dashed lines 
show individual photon lines.
}
\end{figure}

\begin{figure}[htbp]
\includegraphics[width=5in]{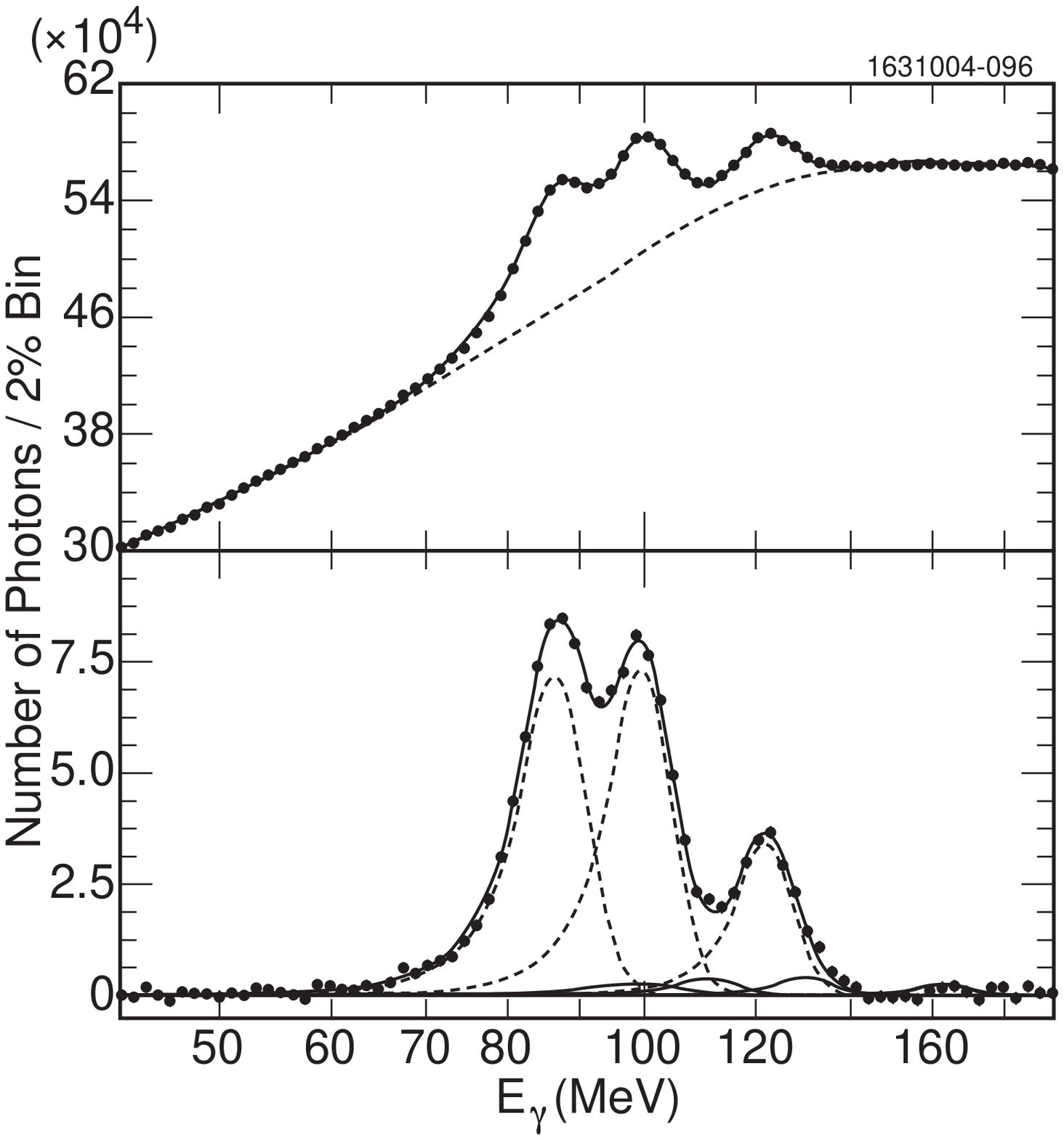}
\caption{
Fit to the $\Y(3S)\to\gamma\chi_{bJ}(2P)$ ($J=2,1,0$) photon
lines in the data.
See caption of Fig.~\ref{fig:2sfit} 
for the description.
Small solid-line peaks 
in the bottom plot
show the $\chi_{bJ}(2P)\to\gamma\Y(1D)$
and $\Y(2S)\to\gamma\chi_{bJ}(1P)$ 
contributions.
\label{fig:3sfit}
}
\end{figure}

The fits for $\Y(2S)\to\gamma\chi_{bJ}(1P)$ 
and for $\Y(3S)\to\gamma\chi_{bJ}(2P)$
photon lines are
shown in Figs.~\ref{fig:2sfit} and \ref{fig:3sfit},
respectively. 
Since natural widths of the $\chi_b$ states are much smaller
than the detector resolution, we represent each photon
line as a Gaussian with an asymmetric low-energy tail
induced by the transverse and longitudinal shower energy
leakage out of the group of crystals used in the
photon energy algorithm. This so-called Crystal Ball line
shape is discussed in more detail elsewhere \cite{psipprl}. 
Shape parameters are varied to estimate 
systematic uncertainties.
The widths of the Gaussian parts
are constrained between the three lines to follow the energy
dependence of the energy resolution predicted by 
Monte Carlo simulation. However, the absolute scale of
the energy resolution is allowed to float in the fit.
The fitted values of this scale factor are
consistent between the $\Y(2S)$, $\Y(3S)$ and $\psi(2S)$ data.

The $\Y(3S)\to\gamma\chi_{bJ}(2P)$
fit includes additional 
Doppler-broadened peaks due to
$\Y(2S)\to\gamma\chi_{bJ}(1P)$ and
$\chi_{bJ}(2P)\to\gamma\Y(1D)$. 
Parameters of these peaks are fixed in the fit from
the other measurements and theoretical 
predictions \cite{RosnerD}.
The systematic errors include variations in the
values of these parameters within the experimental
and theoretical uncertainties.
The obtained photon yields, selection efficiencies
including proper photon angular distributions,
determined branching ratios (${\cal B}$) and
peak energies are given in Table~\ref{tab:e1results}.
Since the statistical errors are very small, the
results are dominated by the systematic uncertainties.

\begin{table}[htbp]
\caption{The results for $\Upsilon(2S)\to\gamma\chi_{bJ}(1P)$ and
$\Upsilon(3S)\to\gamma\chi_{bJ}(2P)$ ($J=0,1,2$) transitions.
The first errors are statistical, the second systematic.
\label{tab:e1results}}
\def\ppmm{\!\pm\!}
\def\1#1#2#3{\multicolumn{#1}{#2}{#3}}
{
\begin{center}
\begin{tabular}{cccc}
\hline\hline
 initial state     & \1{3}{c}{$\Upsilon(2S)$} \\ 
 final state       & $\chi_{b0}(1P)$ & $\chi_{b1}(1P)$ & $\chi_{b2}(1P)$  \\
\hline
 \#\ of $\gamma$'s ($10^3$)        & $198\ppmm6$ & $407\ppmm7$ & $410\ppmm6$ \\
 
 Efficiency ($\%$)    & $57$ & $63$ & $61$ \\
 ${\cal B}$ ($\%$)   & $3.75\ppmm0.12\ppmm0.47$ & $6.93\ppmm0.12\ppmm0.41$ & $7.24\ppmm0.11\ppmm0.40$ \\
 $E_\gamma$ (MeV)    & $162.56\ppmm0.19\ppmm0.42$ & $129.58\ppmm0.09\ppmm0.29$ & $110.58\ppmm0.08\ppmm0.30$ \\
\hline
 initial state     & \1{3}{c}{$\Upsilon(3S)$}  \\ 
 final state       
                   & $\chi_{b0}(2P)$ & $\chi_{b1}(2P)$ & $\chi_{b2}(2P)$  \\
\hline
 \#\ of $\gamma$'s ($10^3$) 
                              & $225\ppmm7$ & $537\ppmm7$ & $568\ppmm6$ \\
 
 Efficiency ($\%$)    
                   & $57$ & $63$ & $61$ \\
 ${\cal B}$ ($\%$)   
                   & $6.77\ppmm0.20\ppmm0.65$ & $14.54\ppmm0.18\ppmm0.73$ & $15.79\ppmm0.17\ppmm0.73$ \\
 $E_\gamma$ (MeV)    
                   & $121.55\ppmm0.16\ppmm0.46$ & $99.15\ppmm0.07\ppmm0.25$  & $86.04\ppmm0.06\ppmm0.27$ \\

\hline\hline
\end{tabular}
\end{center}
\vskip-0.5cm
\quad
}
\end{table}

The largest systematic error in the $E_\gamma$ measurements
is due to the uncertainty in the absolute
calibration of photon energies.
This calibration error was improved
over the one in the
$\psi(2S)\to\gamma\chi_{cJ}(1P)$ 
photon energy measurements \cite{psipprl} by
turning the latter results into recalibration
points, given that the masses of the $\psi(2S)$
and $\chi_{cJ}(1P)$ states are known precisely
from the scans of their resonant cross-sections 
in $e^+e^-$ and $\bar pp$ collisions \cite{PDG}.
The results for photon energies in transitions
to the $\chi_{bJ}(1P)$ and $\chi_{bJ}(2P)$ states
are in good agreement with previous 
measurements \cite{CUSBI}-\cite{CLEOII}.
They are the most precise determinations.
When combined with the measured masses of the
$\Y(2S)$ and $\Y(3S)$ resonances, they allow for a
determination of the masses of the $\chi_{bJ}$ states.
A ratio of the fine mass splitting, 
$r\equiv (M(\chi_{b2})-M(\chi_{b1}))/(M(\chi_{b1})-M(\chi_{b0}))$,
determined from our new
photon energy measurements, gives very similar values for
the $1P$ and $2P$ triplets:
$0.574\pm0.005\pm0.011$ and
$0.584\pm0.006\pm0.013$, respectively.

The branching ratio results for 
$\Y(2S)\to\gamma\chi_{bJ}(1P)$
are also in good agreement with 
previous measurements \cite{CUSBI}, \cite{CB}-\cite{CLEOII} 
and offer improved experimental errors.
However, the branching ratio results for
$\Y(3S)\to\gamma\chi_{bJ}(2P)$
are significantly larger than previous
best measurements by CUSB II \cite{CUSBII} based
on photon samples which were about
a factor of 50 smaller than
utilized in this work.

In the non-relativistic limit, branching
ratios to different members of the same
triplet differ only by the phase-space
factors: $(2J+1) E_\gamma^3$.
We can test this prediction by calculating
the ratio of the branching ratios, divided
by the phase-space factors.
The ratio $(J=2)/(J=1)$
is consistent with unity within the experimental
errors:  $1.009\pm0.016\pm0.077$ 
($0.997\pm0.014\pm0.051$) for $1P$ ($2P$) states.
However, the rate to the spin zero states
appears to be lower: 
$0.822\pm0.020\pm0.063$ ($0.758\pm0.019\pm0.071$) for
the $(J=0)/(J=1)$ $1P$ ($2P$) ratio and
$0.814\pm0.020\pm0.112$ ($0.760\pm0.021\pm0.087$)
for the $(J=0)/(J=2)$ $1P$ ($2P$) ratio.
Suppression of the $J=0$ matrix element was
predicted by theoretical calculations implementing
relativistic corrections \cite{MoxhayRosner}-\cite{McClaryByers}.

\begin{figure}[htbp]
\includegraphics[width=5in]{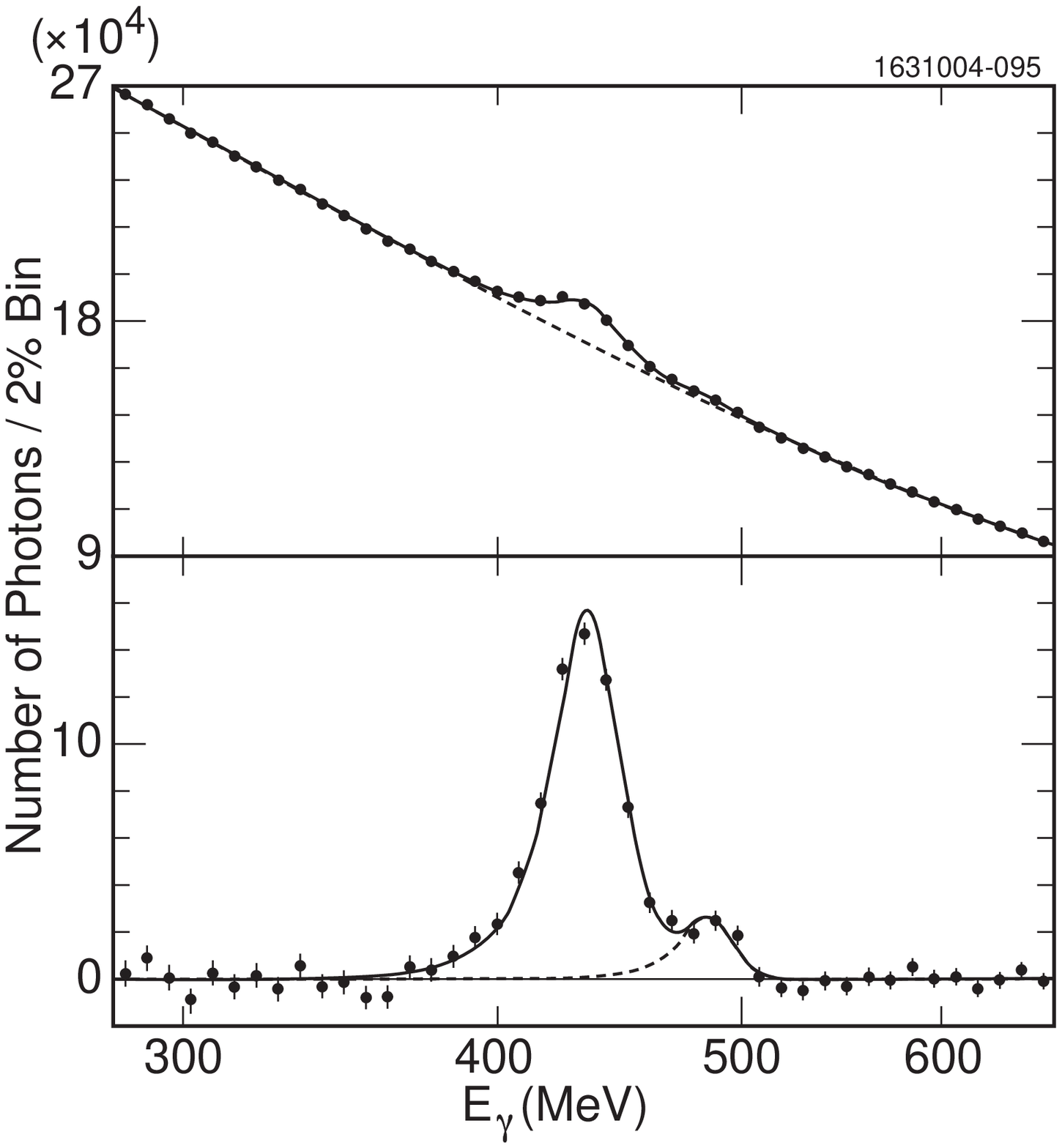}
\caption{
Fit to six photon lines due to
$\Y(3S)\to\gamma\chi_{bJ}(1P)$  and
$\chi_{bJ}(1P)\to\gamma\Y(1S)$ ($J=2,1,0$) 
in the data.
See caption of Fig.~\ref{fig:2sfit} 
for the description.
The dashed line 
in the bottom plot
shows the $\Y(3S)\to\gamma\chi_{b0}(1P)$
contribution by itself.
\label{fig:1p0fit}
}
\end{figure}

The $\Y(2S)$ photon spectrum in Fig.~\ref{fig:all}
shows a peak also at high energies (around 400 MeV) which is
due to $\chi_{bJ}(1P)\to\gamma\Y(1S)$ transitions.
At these high energies our energy resolution is larger than
the fine structure of these states, therefore the 3 peaks
overlap each other. Systematic effects prevent us from
reliably extracting individual line amplitudes.
The similar cascade peaks, $\chi_{bJ}(2P)\to\gamma\Y(2S)$
(around 250 MeV) and $\chi_{bJ}(2P)\to\gamma\Y(1S)$
(around 800 MeV), are also seen in the $\Y(3S)$ spectrum
(Fig.~\ref{fig:all}).

The intermediate high energy peak in the $\Y(3S)$ spectrum
(around 400 MeV) is due to the overlap
of primary, but rare, $\Y(3S)\to\gamma\chi_{bJ}(1P)$ transitions
and cascades $\chi_{bJ}(1P)\to\gamma\Y(1S)$.
The latter are fed not only by the direct transitions 
from the $\Y(3S)$ to the $\chi_{bJ}(1P)$ states, but also by 
photon or hadronic transitions
via the $\Y(2S)$ and $\Y(1D_J)$ states.
One line, $\Y(3S)\to\gamma\chi_{b0}(1P)$, has a photon energy
sufficiently different from the rest of them that its
amplitude can be measured. 
Amplitudes of the other two lines in this triplet,
$\Y(3S)\to\gamma\chi_{b1,2}(1P)$, cannot be reliably 
determined as they overlap each other and the
cascade lines. The photon spectrum with 
$\pi^0$ suppression is used for this portion of the analysis.
The three amplitudes of the photon lines for the direct
transitions, $\Y(3S)\to\gamma\chi_{bJ}(1P)$ ($J=0,1,2$),
are free parameters in the fit. 
They also contribute to the amplitudes
of the cascade lines, 
$\chi_{bJ}(1P)\to\gamma\Y(1S)$.
The other contributions to the latter peaks are
fixed from the measured (for the $\Y(2S)$ route) 
or predicted \cite{RosnerD}
(for the $\Y(1D_J)$ route) branching ratios. 
All photon energies are fixed to the known masses
of the states involved.
The fit is displayed in Fig.~\ref{fig:1p0fit}.
The number of observed $\Y(3S)\to\gamma\chi_{b0}(1P)$ events is
$8,700\pm1,100$ with a selection efficiency of $49\%$.
The corresponding branching ratio is
${\cal B}(\Y(3S)\to\gamma\chi_{b0}(1P))$
$=(0.30\pm0.04\pm0.10)\%$. 
This is the first determination of this rate.
While various potential model calculations
roughly agree with each other and our results
for the rates of $\Y(2S)\to\gamma\chi_{bJ}(1P)$
and $\Y(3S)\to\gamma\chi_{bJ}(2P)$ transitions,
the predictions for the 
$\Y(3S)\to\gamma\chi_{b0}(1P)$ rate
vary substantially and most of them are inconsistent with
our measurement, as illustrated by the following
sample of theoretical predictions \cite{CLEOBmm}:
$0.006\%$ 
\cite{Gupta},
$0.05\%$ 
\cite{Fulcher},
$0.12\%$ 
\cite{MoxhayRosner},
$0.13\%$ 
\cite{efg04},
$0.21\%$ 
\cite{Daghighian},
$0.56\%$ 
\cite{Grotch} and
$0.74\%$ 
\cite{lah03}.

We have also analyzed our data for possible photon
peaks due to hindered M1 transitions to singlet
$\eta_b(1S)$ and $\eta_b(2S)$ states.
Since the masses of these states are not known
experimentally, we search for these transitions
at photon energies corresponding to
theoretical predictions \cite{RosnerM1}:
$\eta_b(1S)$ ($\eta_b(2S)$) mass $35-110$ ($20-45$) MeV
below the $\Y(1S)$ ($\Y(2S)$) mass.
No evidence for such transitions is found.
We set the following 90\%\ upper limits:
${\cal B}(\Y(3S)\to\gamma\eta_b(1S))<4.3\times10^{-4}$,
${\cal B}(\Y(3S)\to\gamma\eta_b(2S))<6.2\times10^{-4}$
and
${\cal B}(\Y(2S)\to\gamma\eta_b(1S))<5.1\times10^{-4}$.
These limits are a factor 5-7 lower than 
the branching ratio we measured 
for the hindered M1 transition in the $c\bar c$ system,
$\psi(2S)\to\gamma\eta_c(1S)$ \cite{psipprl,CBcc}. 
The limit on ${\cal B}(\Y(3S)\to\gamma\eta_b(1S))$ 
appears to provide the tightest constraint on theoretical
predictions, as illustrated by the following
sample of theoretical predictions \cite{CLEOBmm}:
$29\times10^{-4}$ 
\cite{zb83,RosnerM1},
$(14-32)\times10^{-4}$ 
\cite{GI85,RosnerM1},
$5.2\times10^{-4}$ 
\cite{efg04} and
$3.6\times10^{-4}$ 
\cite{lah03}.
Only the last prediction is consistent with our
upper limit.

In summary, we have reported improved photon energy and branching
ratio measurements for the E1 transitions:
$\Upsilon(2S)\to\gamma\chi_{bJ}(1P)$ and
$\Upsilon(3S)\to\gamma\chi_{bJ}(2P)$ ($J=0,1,2$).
Our results are consistent with previous measurements,
except for the branching ratios for the latter transitions
which are found to be significantly higher.
We have measured the rare E1 transition
$\Upsilon(3S)\to\gamma\chi_{b0}(1P)$ 
for the first time.
No evidence for hindered M1 transitions to $\eta_b(1S)$ and
$\eta_b(2S)$ states is found in our data, against 
the expectations from some theoretical calculations.

We gratefully acknowledge the effort of the CESR staff in providing us with
excellent luminosity and running conditions.
This work was supported by 
the National Science Foundation and
the U.S. Department of Energy.


\begin{thebibliography}{99}

\bibitem{LatticeQCD}
C.T.H.~Davies,
Phys.\ Rev.\ Lett.\ {\bf 92}, 022001 (2004).

\bibitem{PotentialModels}
For reviews see e.g.\ 
D.~Besson, T.~Skwarnicki,
Annu.\ Rev.\ Nucl.\ Part.\ Sci.\ {\bf 43}, 333 (1993);
E.J.~Eichten, C.~Quigg, 
Phys.\ Rev.\ D {\bf 49}, 5845 (1994).

\bibitem{CUSBI}
CUSB, 
K.~Han {\it et al.},
Phys.\ Rev.\ Lett.\  {\bf 49}, 1612 (1982);
C.~Klopfenstein {\it et al.},
Phys.\ Rev.\ Lett.\  {\bf 51}, 160 (1983);

\bibitem{CUSBII}
CUSB II, 
U.~Heintz {\it et al.}
Phys.\ Rev.\ D {\bf 46} 1928 (1992).

\bibitem{CB}
Crystal Ball, 
R.~Nernst {\it et al.},
Phys.\ Rev.\ Lett.\ {\bf 54}, 2195 (1985).

\bibitem{ARGUS}
ARGUS, 
H.~Albrecht \etal, 
Phys.\ Lett.\ B{\bf 160}, 331 (1985).

\bibitem{CLEOI}
CLEO, 
P.~Haas \etal, 
Phys.\ Rev.\ Lett.\ {\bf 52}, 799 (1984).

\bibitem{CLEOII}
CLEO II, 
R.~Morrison \etal, 
Phys.\ Rev.\ Lett.\ {\bf 67}, 1696 (1991);
K.W.~Edwards  \etal,
Phys.\ Rev.\ D {\bf 59} 032003 (1999).

\bibitem{psipprl}
CLEO III, 
S.~B.~Athar \etal,
CLNS 04/1886, CLEO 04-10,
arXiv:hep-ex/0408133, 
{\it submitted to PRD}.

\bibitem{CLEOIIdetector}
CLEO II, 
Y.~Kubota {\it et~al.},
Nucl.\ Instr.\ Meth.\ A {\bf 320}, 66 (1992).

\bibitem{CLEOIIIDR} 
D.~Peterson {\it et~al.}, 
Nucl.\ Instr.\ Meth.\ A {\bf 478}, 142 (2002).


\bibitem{PDG}
PDG, 
S.~Eidelman {\it et al.},
Phys.\ Lett.\ B {\bf 592}, 1 (2004).

\bibitem{MoxhayRosner}
P.~Moxhay, J.~L.~Rosner, Phys.\ Rev.\ D {\bf 28}, 1132 (1983).

\bibitem{McClaryByers}
R.~McClary, N.~Byers, Phys.\ Rev.\ D {\bf 28}, 1692 (1983).
 
\bibitem{RosnerD}
S.~Godfrey, J.L.~Rosner, 
Phys.\ Rev.\ D {\bf 64}, 097501 (2001);
{\it erratum} D {\bf 66} 059902 (2002);
W.~Kwong, J.~L.~Rosner, 
Phys.\ Rev.\ D {\bf 38}, 279 (1988).

\bibitem{CLEOBmm}
These theoretical predictions were rescaled to
the recent value of the total width of the
$\Y(3S)$ resonance - see
CLEO III, 
G.S. Adams {\it et al.},
CLNS 04-1887, arXiv:hep-ex/0409027, 
{\it submitted to PRL}.

\bibitem{Gupta}
S.~N.~Gupta, S.~F.~Radford, W.~W.~Repko, 
Phys.\ Rev.\ D {\bf 30}, 2424 (1984).

\bibitem{Fulcher}
L.~P.~Fulcher, Phys.\ Rev.\  D {\bf 42}, 2337 (1990).

\bibitem{efg04} 
D. Ebert, R. Faustov, V. Galkin, 
Phys. Rev. D {\bf 67}, 014027 (2003).

\bibitem{Daghighian}
F.~Daghighian, D.~Silverman,
Phys.\ Rev.\ D {\bf 36} 3401 (1987). 

\bibitem{Grotch}
H.~Grotch, D.~A.~Owen, K.~J.~Sebastian, 
Phys.\ Rev.\ D {\bf 30}, 1924 (1984).

\bibitem{lah03}
T.A. L\"{a}hde, Nucl.\ Phys. A{\bf 714}, 183 (2003).

\bibitem{RosnerM1}
S.~Godfrey, J.L.~Rosner,
Phys.\ Rev.\ D {\bf 64} 074011 (2001); 
erratum-ibid. D {\bf 65} 039901 (2002)
and references therein.

\bibitem{CBcc}
Crystal Ball, 
J.E.~Gaiser {\it et al.}, 
Phys.\  Rev.\ D {\bf 34}, 711 (1986).

\bibitem{zb83}
V. Zambetakis, N. Byers, 
Phys.\ Rev.\ D {\bf 28}, 2908 (1983).

\bibitem{GI85} 
S.~Godfrey, N.~Isgur, 
Phys.\ Rev.\ D {\bf 32}, 189 (1985).

\end{thebibliography}
\end{document}